\documentclass[manuscript,screen]{acmart}
\usepackage{subfig}
\usepackage{graphicx}
\AtBeginDocument{%
  \providecommand\BibTeX{{%
    \normalfont B\kern-0.5em{\scshape i\kern-0.25em b}\kern-0.8em\TeX}}}

\acmConference[WSDM IRS Workshop]{2023 WSDM Workshop on Interactive Recommender Systems}{2023}{Singapore}
\begin{document}

\setcopyright{none}
\settopmatter{printacmref=false} 
\renewcommand\footnotetextcopyrightpermission[1]{} 

\title{Dynamic Adaptation of User Preferences and Results in a Destination Recommender System}

\author{Asal Nesar Noubari}
\email{asal.nesar@tum.de}
\affiliation{%
  \institution{Technical University of Munich}
  \streetaddress{Boltzmannstr. 3}
  \city{Garching bei München}
  \postcode{85748}
  \country{Germany}
}

\author{Wolfgang W{\"o}rndl}
\email{woerndl@cit.tum.de}
\affiliation{%
  \institution{Technical University of Munich}
  \streetaddress{Boltzmannstr. 3}
  \city{Garching bei München}
  \postcode{85748}
  \country{Germany}
}


\begin{abstract}
    Studying human factors has gained a lot of interest in recommender systems research recently. User experience plays a vital role in tourism recommender systems since user satisfaction is the main factor that guarantees the success of such recommender systems. In this work, we have designed and implemented a destination recommender system in which the recommendations adapt instantly based on the user preferences. The recommendations can be explored on a world map with additional information. This interface addresses common visualization challenges in recommender systems, such as transparency, justification, controllability, explorability, the cold-start problem, and context awareness. We have conducted a user study to evaluate different aspects of this recommender system from the users' perspective.
\end{abstract}



\maketitle
\emph{In 2023 WSDM Workshop on Interactive Recommender Systems, Singapore, March 3rd, 2023}


\section{Introduction} 
Nowadays, travelers use various online services and recommender systems to plan their trips. Recommender systems allow users to deal with data overload and make better decisions in a personalized way \cite{jannach2010recommender}. Academic research often focuses on recommendation algorithms to generate more accurate recommendations. Nevertheless, user experience is a crucial factor that guarantees user satisfaction and, thus, the success of the recommender system \cite{konstan2012recommender}. In tourism, the recommender system needs to be highly interactive and context-aware \cite{jannach2020interactive}.

Choosing a destination is a main part of travel arrangements. Users have different expectations, budgets, and needs when traveling. Travelers should be able to revise their preferences and instantly get recommendations based on their input. Furthermore, transparency in such recommender systems plays a vital role in gaining users' trust. Even though systems with this application exist, usually, the destination recommenders lack interactivity, context awareness, and transparency. These factors have an immense impact on user experience and they can help users in their decision-making process.

We have designed and implemented a destination recommender system called \emph{Destination Finder} \cite{thesis}. The application is highly interactive and recommendations adapt instantly when user preferences are modified. In this paper, we first explain the background of our work including visualization challenges and related work. Then, we describe the \emph{Destination Finder} application and the ideas behind the user interface elements in Section 3. We have conducted a user study to test the usability of our approach, the setup and results will be presented in Section 4, before concluding the paper with a brief summary and outlook.

\newpage

\section{Background} 
\subsection{Visualization Challenges}
Recommender systems pose challenges concerning visualization as follows:

\begin{enumerate}
\item \textbf{Transparency} of the recommendation techniques: recommender systems often have a black-box nature, but the user needs to know how the system is working to be able to trust its recommendations.
\item  \textbf{Justification} of the reason behind the recommendations: justification (or explanation) helps the user understanding why an item or a list of items was recommended. Many recommender approaches include some form of explanation to justify their results.
\item \textbf{Controllability} as user control over the user model: allows the user to be involved directly with the recommender systems’ results. 
\item \textbf{Explorability} of the item space: Explorability can be defined as providing the user with the visualizations to browse the entire information space and not just the recommended items. 
\item The \textbf{cold-start problem}: learning new users’ preferences is always challenging in recommender systems. Different ways have been proposed to elicit user preferences with minimal effort. 
\item Acquiring contextual information and \textbf{context-awareness}: incorporating contextual information into the recommendation process has gained a lot of interest over the past years. For example, recommender systems on mobile devices use the user’s location, current time, or the weather to recommend appropriate items in the vicinity.
\end{enumerate}

\subsection{Related Work}
 \cite{jannach2020interactive} reviews interactive and context-aware systems in tourism, where recommender systems play an important role. These systems include the recommendation of destinations, hotels, events, restaurants, or points-of-interests in general. Thereby, most research approach focuses on recommendation algorithms and
offline experimentation. In the context of tourism, traditional approaches cannot be directly applied because, firstly, interactive acquisition of users' needs and preferences is required \cite{jannach2020interactive}. Secondly,  the suitability of a recommendation depends on the given context of the user such as the current location \cite{jannach2020interactive}. \cite{ricci2022recommender} discusses important dimensions for recommender systems in the scenario of travel and tourism. 

\cite{keck2018exploring} is an approach in a different domain that influenced our design. Their presented application allows for choosing a movie for one person and a group of people with different tastes. The system recommends a movie based on the whole group’s taste but also visualizes individual preferences. This is done by first showing clusters of items that are zoomable and explorable for the users. In addition, circles with different colors are used to represent the individual preferences and the weights of a user in contributing to the overall score of the item.

The \emph{PARIS} \cite{jin2016go} recommender system allows the user to control the recommendation process by adjusting their profile with input controls such as drop-down lists and checkboxes. This system works with the user's characteristics such as age, gender, and personality, which are received from the user input, and describes which data is used in which step to select the recommended items. The movie recommender by Loepp et al. (2014) \cite{loepp2014choice} allows users to iteratively choose between two movie choices to elicit their preference when new to the system.

 Map-based interfaces can resolve some of the mentioned visualization challenges including transparency, explorability, and context-awareness \cite{kunkel2019map, averjanova2008map}. \cite{kunkel2019map} proposes a method to make recommendations more comprehensible and controllable even in application areas with items that are not location-based (e.g. movies). The item space is visualized using a map-like interface to improve the users' overview of the domain. The user can then directly interact with the map interface to adjust preferences and thus refine recommendations. Their conducted user studies with prototypes showed good potential of the proposed method to increase overview, transparency and control in recommender systems.

Many existing recommender applications address some of the presented challenges well. For example, \emph{whichbook} recommends books based on mood and emotion\footnote{https://www.whichbook.net/mood-emotion}. The interesting part is that once you change the sliders on the left, the recommendation on the right adapts immediately without further user input. So the application offers good controllability, mitigates the cold start problem, and is context-aware. However, other challenges are not addressed.

\begin{figure}[!htb]
  \centering
  \includegraphics[width=1.0\textwidth]{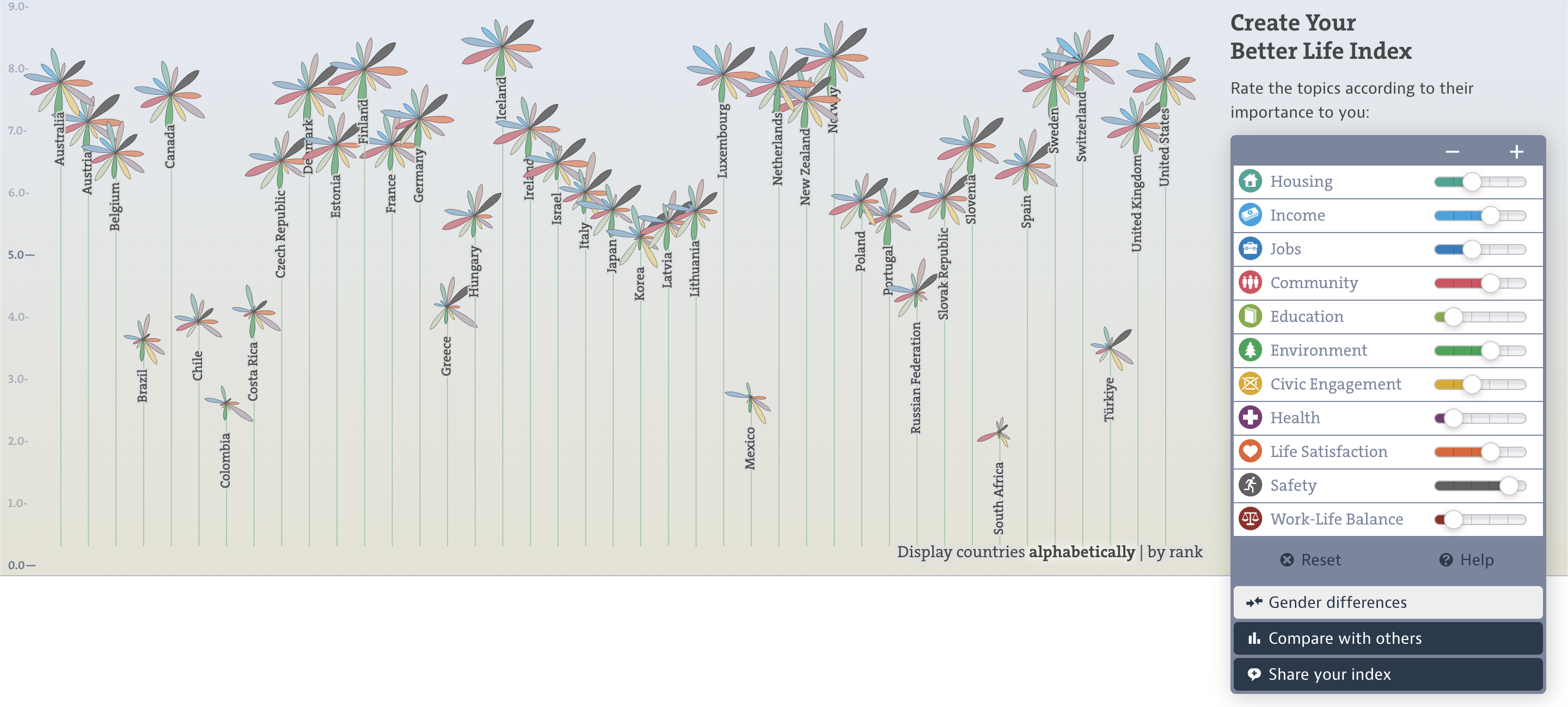}
  \caption{OECD Better Life Index} \label{fig:oecd}
\end{figure}

\emph{OECD Better Life Index}\footnote{https://www.oecdbetterlifeindex.org} is an application that calculates the life index of all countries based on the user's preferences and priorities (Fig. \ref{fig:oecd}). The user should rate the above topics based on their importance to them. They can instantly see the effect of their preference customization on the results. The results demonstrate a set of graphs for each country, showing the score of each topic for that specific country and how close those scores are to the user's ratings. These topics are differentiated by the colors which match the sliders that contain users' ratings. However, the user interface is not intuitive and lacks transparency.

\newpage

\section{The Destination Finder Web Application} 
Our application \emph{Destination Finder} allows for specifying travel budget, intended duration, and user preferences for activities and visualizes recommended destinations on a map \footnote{The source code of the application can be found here:  https://github.com/asalnesar/destination-finder}.

\subsection{Main Elements of the User Interface} 

\begin{figure}[!htb]
  \centering
  \includegraphics[width=1.0\textwidth]{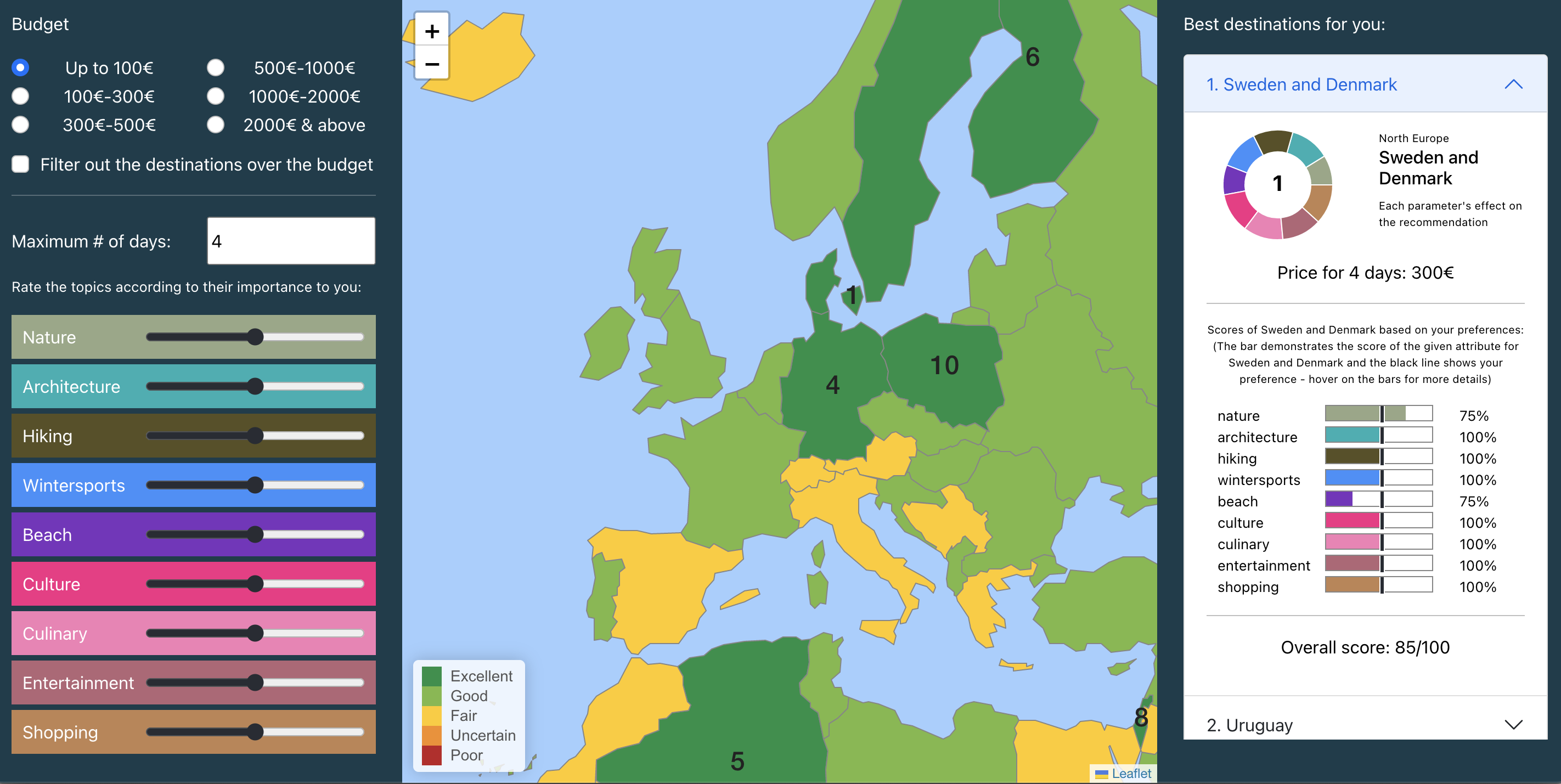}
  \caption{Main Interface of \emph{Destination Finder}} \label{fig:dest-finder-main}
\end{figure}

Figure \ref{fig:dest-finder-main} shows the main interface of the \emph{Destination Finder} application \footnote{A demo of the application can be accessed here: https://destination-finder.netlify.app/}. The application has three main panels: the preference customization panel, the map panel, and the result panel.
  
The system uses the user input with the preference customization panel (Fig. \ref{fig:dest-finder-main}, left) to instantly revise the recommendations based on the region attributes and user preference, and reflect them on the map panel as well as present the 10 best matching results in the results panel on the right. The preference customization panel allows inputting budget, and activity preferences. Users can specify a budget level using radio buttons, the maximum number of days and whether destinations over the budget limit are eliminated from consideration. Users can also indicate their preferences about nine activities or topics with sliders: nature, architecture, hiking, winter sports, beach, culture, culinary, entertainment, and shopping.

The map panel (Fig. \ref{fig:dest-finder-main}, center) illustrates how well a travel region matches the user preferences. Our region model is not solely based on country borders. Bigger countries such as the United States of America, Canada or Australia are divided into several regions because of the attribute variation in different locations within them. Smaller countries are merged, e.g. the Netherlands, Belgium and Luxembourg form one travel region. The map panel has all the standard user controls, the user can move the map or zoom in or out using the handles available on the top left of the map or by scrolling. Regions are scored based on the similarity of their attributes to the user’s preferences. The regions’ colors indicate their score range. Green means that the region has an excellent score, light green means the region has a good score, yellow means a fair score, orange regions have an uncertain score, and regions with red background color have a poor score. These categories are visible on the legend located on the bottom left of the map panel. Furthermore, the 10 best matching destinations based on the user’s preferences are labeled with their ranks, so they are more distinct on the map. Clicking on the map opens a popup window with more information about the respective travel regions. Our map panel visualizes the whole item space for good explorability.

\begin{figure}[!htbp]
  \centering
  \subfloat[][Winter Sports and Greece]{\includegraphics[width=0.5\textwidth]{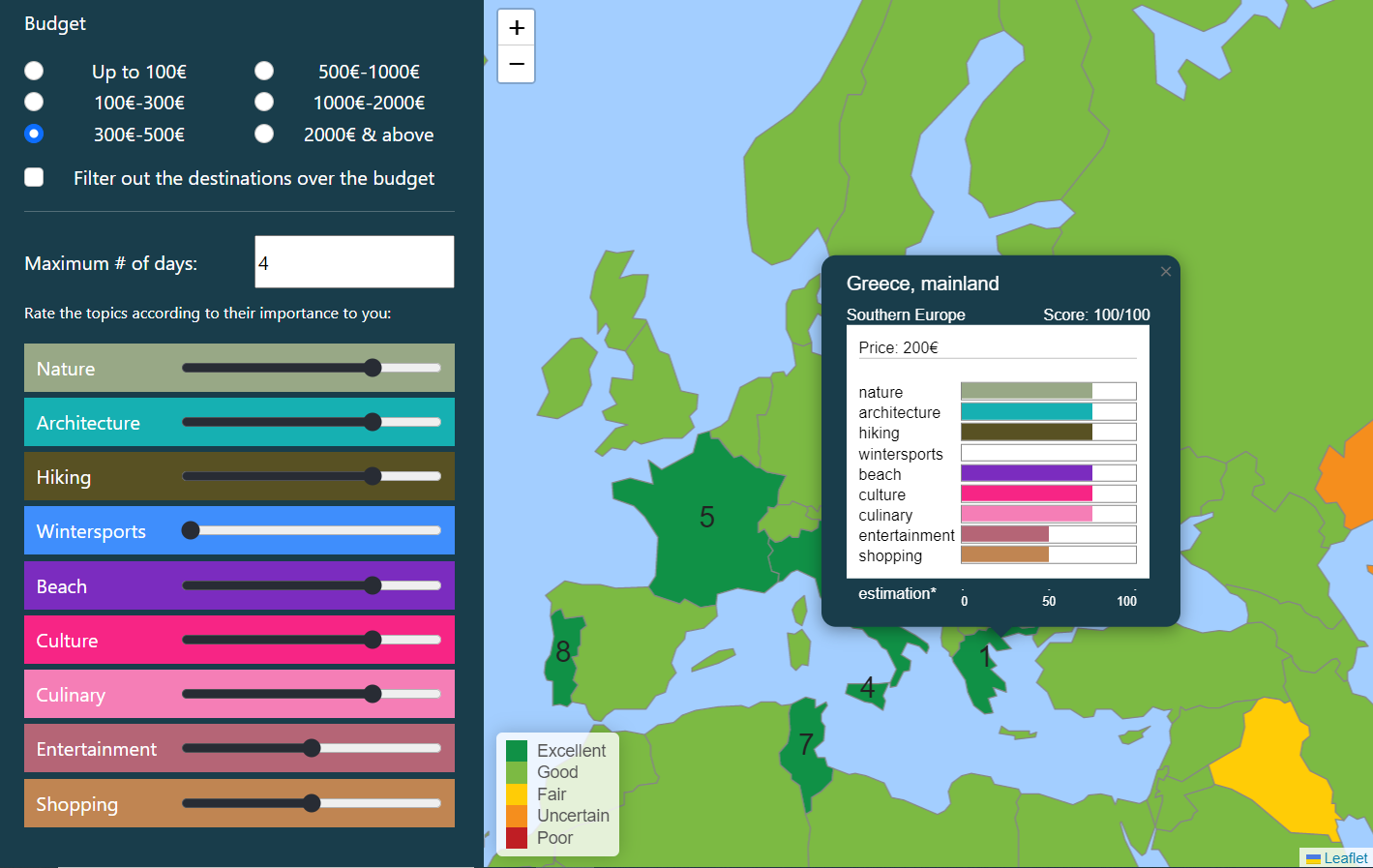}\label{fig:preferences1}}
  \subfloat[][Winter Sports and France]{\includegraphics[width=0.5\textwidth]{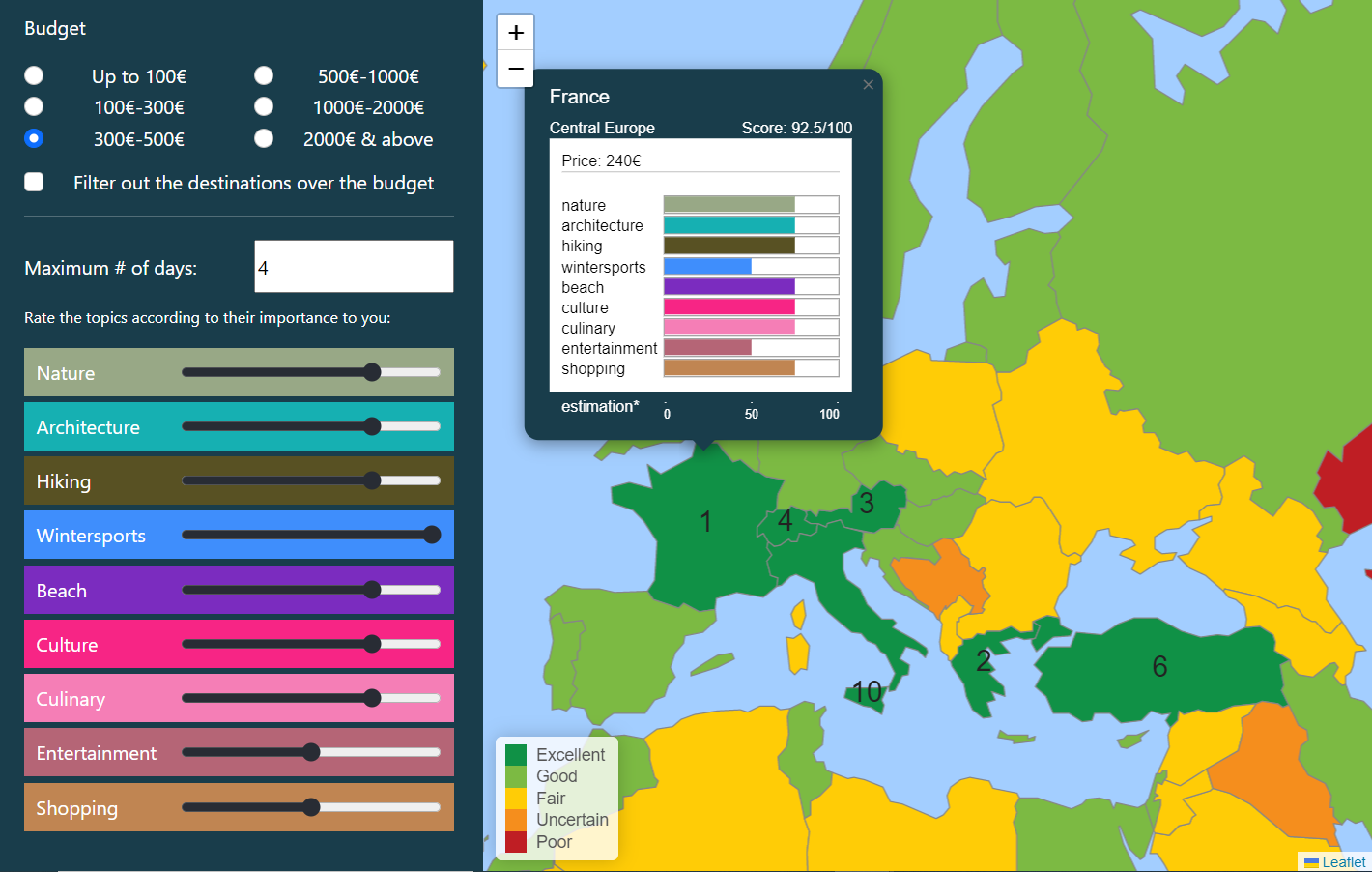}\label{fig:preferences2}}
  \caption{Impact of Preference Change}
  \label{fig:destination-finder-pref}
\end{figure}

Note that changing the user preferences instantly change the recommendation scores and map visualization. For example in \ref{fig:destination-finder-pref}, Greece is the best matching region according to the user preferences. When moving the slider for "winter sports" from the left (not important) to the right (important), France replaces Greece as the best matching region and several other regions have changed score ranges as indicated by the colors on the map.

\begin{figure}[!htbp]
  \centering
  \subfloat[][List with Best Matching Destinations]{\includegraphics[height=0.4\textheight]{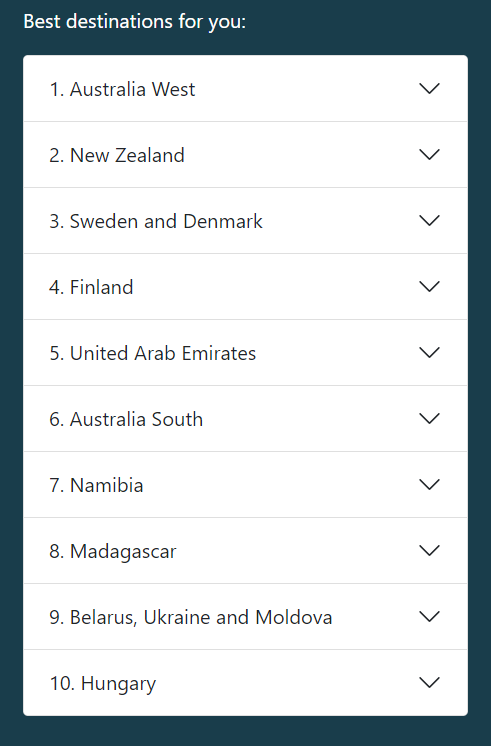}\label{fig:results1}}
  \subfloat[][Result Information Panel Expanded]{\includegraphics[height=0.4\textheight]{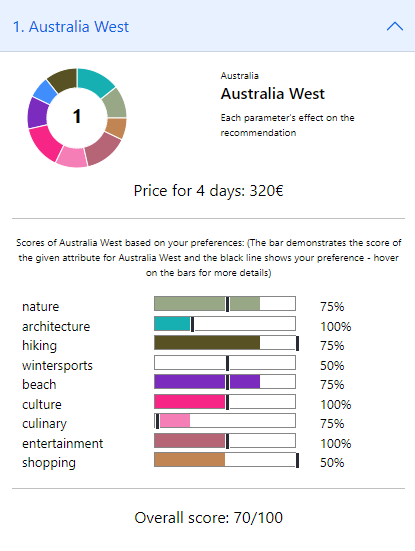}\label{fig:results2}}
  \caption{Results Panel}
  \label{fig:destination-finder-results}
\end{figure}

The results panel (Fig. \ref{fig:dest-finder-main}, left) shows the 10 best matching regions in a list (Fig. \ref{fig:results1}). The items can be expanded to present more information about that region (Fig. \ref{fig:results2}).

The pie chart displayed in Figure \ref{fig:results2} shows the effect of each attribute on the destination's success in making it to the 10 best matching recommendations. For example, in figure \ref{fig:results2}, the user can observe that the region is recommended because its architecture, culture and entertainment matches their corresponding preferences. The pie chart's colors are the same as the attribute sliders' backgrounds in the preference customization panel (Fig. \ref{fig:dest-finder-main}, left). 

The bar charts displayed in Figure \ref{fig:results2} provide the most important information about the recommendation. Each attribute has a bar chart with the same color as the attribute sliders' backgrounds in the preference customization panel. Furthermore, each bar chart has a benchmark (a vertical black line) that indicates the user's preference for that attribute. This value is the same as the value of the slider of that specific attribute, which was customized by the user in the preference customization panel. So the users can easily compare the region attributes with their preferences. For example in Figure \ref{fig:results2}, the region Australia West has a score of 75/100 for nature (first bar chart), while the user has set the corresponding slider to the middle value 50/100. The application provides a tooltip explaining the scoring when hovering over a bar chart. At the bottom of the panel, the overall score of the recommended country is displayed. Thus, the application offers good justification and the users know why a destination was recommended.

\subsection{Architecture, Data and Recommendation Algorithm}
The software is based on the Model-View-Controller (MVC) architectural pattern. MVC divides the application into three corresponding parts and increases the reusability and maintainability of the software. For the model, the system uses preferences as specified in the user interface, data about travel regions and also \emph{GeoJSON} information to implement the interactive map. \emph{GeoJSON} is a format for encoding various geographic data structures that uses a geographic coordinate reference system. This data is static and does not change in the application. The \emph{Destination Finder} web application is a single-page application with some interfaces for receiving user input, an interactive map panel, and a list containing recommendations and detailed scoring and information. This is the View of the MVC pattern. The Controller triggers whenever a preference is changed by the user. It then calculates the recommendation scores for all regions based on the user input and the regions’ information and modifies the view based on the new results. 

Currently, the application does not have a back-end server part. The project’s focus is on the user interface and to recommend items instantly. Therefore, the whole application logic has been implemented on the client side to improve performance and interactivity. The front-end application (View) has been implemented with \emph{ReactJS}, an open-source JavaScript library built by Facebook.

The application utilizes data about travel regions and their suitability for travel attributes, along with price level estimates and other characteristics. This data was gathered from different sources for earlier research \cite{worndl2017web}. As the main focus in this work was on the dynamic adaptation of user control in the user interface, a simple linear similarity formula is used to calculate a score of each destination region. The users rate activity attributes based on their importance to them between 0 to 100. On the other hand, each region has a score for each of these attributes. The algorithm calculates the difference between both numbers and the average difference over all attributes. For the budget, the data model has basic estimates for minimum costs, this number is compared to the user preferences with regard to total budget and minimum number of days. If the estimated minimum costs are below the budget, this attribute is assumed to be fulfilled and included in the score calculation. The user has the option to specific "filter out the destinations over the budget" in the user interface. If this is selected, the score of all regions above the budget is set to 0. Finally, all regions are ranked according to the calculated scores.

\section{User Study} 
We have evaluated our application in a preliminary user study.

\subsection{Methodology and Setup}
The evaluation was based on the \emph{ResQue} framework \cite{pu2011user} using a questionnaire. \emph{ResQue} is an evaluation framework that measures a recommender system's quality from a user's perspective through criteria such as the recommendations’ quality but also the system’s usability, and the users’ intention to reuse the system. Our adapted questionnaire included 19 questions on a five-point Likert Scale in categories such as:
\begin{itemize}
    \item Perceived Ease of Use (e.g. "I became familiar with the recommender system very quickly, and I found it easy to use.")
    \item Interface Adequacy (e.g. "I found it intuitive to modify my preferences in the recommender using the sliders.")
    \item Perceived Usefulness e.g. "The impact of the number of staying days on recommendations was clear and easy to understand.")
    \item Interaction Adequacy (e.g. "The available travel topics are sufficient for me to make a decision about my travel destination.")
    \item Control, Explanation \& Transparency (e.g. "The information provided for the recommended destinations helped me understand why the items were recommended to me.")
\end{itemize}

The participants were asked to use the \emph{Destination Finder} application and answer the survey questions. An invitation was sent on platforms like a university forum, Linkedin, and university groups on Facebook and Telegram, so almost 93\% of the participants were between ages 20-40 and familiar with similar web applications. Overall, 70 users participated in the study.

\subsection{Results Overview}
The user study showed that most participants (90\%) were generally satisfied with the recommender system and found the \emph{Destination Finder} application helpful for deciding on their future travel destinations\footnote{Due to space constraints in the paper, we can only summarize the study outcomes, more detailed results can be found in \cite{thesis}.}. The survey responses also indicated that the \emph{Destination Finder} application was easy to use from the user's point of view. 97\% of the participants became familiar with the application quickly. More than 85\% stated that the recommendations matched their preferences.

Furthermore, most participants felt in control while using the \emph{Destination Finder} application, and they noticed how their input instantly changed the recommendations. This result means that the implemented recommender system could resolve the controllability in visualization challenges. Most participants found the interaction with the recommender system adequate to make decisions. More than 87\% of the participants stated that the information provided helped them understand why the items were recommended to them. This result shows that the \emph{Destination Finder} addressed the visualization challenges of transparency and justification. More than 80\% stated that the provided information was sufficient for them to decide on their travel destination, while only a few complained about the insufficiency of data or information overload.

Most of the user interface pasts were perceived as positive with regards to interface adequacy. The pie chart in the recommendation panel (Figure \ref{fig:results2}) was the only interface element with significant negative or neutral responses. Responses indicated that the design of the filtering over-budget destinations should improve to some extent to be more intuitive. Similarly, the evaluation of perceived usefulness also showed the users found the bar charts in the recommendation panel useful; however, they were not sure about the usefulness of the pie chart.

We also asked about the users' favorite features of the \emph{Destination Finder} application. The option "Instantly seeing the impact of my choices in the map and the results panel" came in first with 81.4\% of the votes, followed by "Having control over the recommendations by customizing my preferences" with 67.1\%. In contrast, the options "Having the result displayed on a map" and "Transparency as to why and how the destinations are ranked" reached 64.3\% and 31.4\%, respectively.

\section{Conclusion} 
In this paper, we have presented the \emph{Destination Finder} application which allows for interactive elicitation and refinement of user preferences for the scenario of finding a destination to travel to. Results are visualized on a map and color codes are used to illustrate the suitability of a region in comparison to the user's request. An important feature of the application is that the visualization of the item space and its suitability to the user preferences adapts dynamically and instantly when preferences are refined. We have conducted a user study to evaluate various aspects of the system, including interface adequacy, explanation and transparency, perceived ease of use, and perceived usefulness. Results indicated that our approach was successful and thus contributing to transparency, controllability, explanation, and explorability.

Future work could involve the improvement of the underlying data, especially concerning prices and cost. Since most people travel in groups, the application could also be extended to support group recommendations. Members of the group would elicit their preferences individually and different aggregation strategies can be tested to generate reasonable recommendations for the whole group. Group recommendations in this scenario pose interesting visualization challenges, e.g. showing each member’s effect on generating that recommendation. To support group recommendations, our application would need a back-end part to store and manage the information from multiple clients.

Other ideas for future work are to make the user interface more configurable by for example offering a novice or an export mode with different levels of detail and control. In addition, context-awareness can be improved by considering the user's location and estimated transportation costs to destinations, or by taking the destination’s weather during the desired travel period into account.

\newpage


\bibliographystyle{ACM-Reference-Format}
\bibliography{paper-irs-wsdm23}


\end{document}